\documentclass[prd,showpacs,showkeys,amsmath,amssymb]{revtex4}%
\usepackage[dvips]{graphics}
\usepackage{array}

\begin{document}


\title{Lattice calculation of gluon screening masses}

\author{A.~Nakamura,}
\affiliation{Research Institute for Information Science and Education,
 Hiroshima University, Higashi-Hiroshima 739-8521, Japan}
\author{T.~Saito}
\affiliation{Research Institute for Information Science and Education,
 Hiroshima University, Higashi-Hiroshima 739-8521, Japan}
\author{S.~Sakai}
\affiliation{Faculty of Education, Yamagata University,
 Yamagata 990-8560, Japan}


\begin{abstract}
We study $SU(3)$ gluon electric and magnetic masses 
at finite temperatures using quenched lattice QCD
on a $20^2 \times 32 \times 6$ lattice.
We focus on temperature regions between $T=T_c$ and $6T_c$, which are
realized in BNL Relativistic Heavy Ion Collider and CERN Large Hadron 
Collider experiments. 
Stochastic quantization with a gauge-fixing term 
is employed to calculate gluon propagators.
The temperature dependence of the electric
mass is found to be consistent with the hard-thermal-loop perturbation, 
and the magnetic mass has finite values in the temperature region
of interest. 
Both screening masses have little gauge parameter dependence.
The behavior of the gluon propagators 
is very different in confinement/deconfinement physics.
The short distance magnetic part behaves 
like a confined propagator even in the deconfinement phase.
A simulation with a larger lattice, $32^2 \times 48 \times 6 $,
shows that the magnetic mass has a stronger finite size effect
than the electric mass. 
\end{abstract}




\maketitle

\section{Introduction}

One of the most interesting features of QCD (quantum chromodynamics) 
is the transition from the confinement to the deconfinement phase.
In this new state of QCD, quarks and gluons confined in the hadron 
at zero temperature move freely
when the system reaches a sufficiently high temperature.
The quark gluon plasma (QGP) was realized at high temperature 
in the early universe, 
and is expected to be produced in heavy-ion collision experiments at
the CERN Super Proton Synchrotron (SPS), BNL Relativistic Heavy Ion 
Collider (RHIC), and CERN Large Hadron Collider (LHC).
Thus it is an urgent task to accumulate 
theoretical knowledge about the QGP.

The massless gluon in the QGP medium 
is changed into a dressed massive gluon after quantum corrections.
The screening effect is characterized by a mass pole
of the propagator and is closely related to thermal QCD phenomenology.
One example is a screened heavy quark potential,
which is frequently discussed in relation to
 $J/\psi$ or $\Upsilon$ suppression.
For calculations of jet quenching,
which might be a fingerprint of a QGP, a model including the electric 
and magnetic masses has been proposed \cite{wang}.
A nonperturbative quantitative study in the vicinity of $T_c$  
and up to several times $T_c$ is
 of great importance for understanding QGP physics.

The thermal field theory \cite{Gross,Kapusta,Bellac}
is the most basic method of studying the QGP
and has provided many informative observations.
At zero temperature,
many calculations based on perturbative QCD have described 
experiment results \cite{Muta}
and there is no doubt that QCD is a theory of strong interaction.
It is natural to employ the perturbative approach to thermal QCD.
Because of the asymptotic freedom at high temperature,
the coupling constant is expected to
become small enough to carry out the perturbations.
In such a high energy state, quarks and gluons must behave 
as an ideal gas; yet this simple consideration
is spoiled by an infrared divergence \cite{Linde},
which is known to bear a hierarchy on the energy scale 
in the QGP system \cite{Kapusta,Bellac}.
One usually defines $1/T$ as a perturbative length scale,
and furthermore $1/gT$ must be introduced as an electric (Debye) scale, 
whose influence appears as a Yukawa-type potential
rather than a Coulomb-like one, and $1/g^2T$ as a magnetic scale.
In QCD, the magnetic mass, which cannot be accessed perturbatively,
 acts as a cutoff factor in the infrared problem and 
consequently becomes an essential element of thermal QCD.

The QCD coupling constant strength near the critical temperature $T_c$ 
is still of the order of 1. 
Therefore perturbation theory is not applicable.
However, these regions are currently being investigated with great interest
in much theoretical and experimental research.
After hard-thermal-loop (HTL) resummation 
was consistently formulated by Braaten and Pisarski,
there were several improvements \cite{BP,Niegawa}
for unsolved problems.
Comparison of this method with lattice numerical data has been reported;
 one-loop HTL calculations of the free energy of a QGP
are in good agreement with the lattice numerical result \cite{Andersen,Andersen2}.
However, a recent two-loop HTL calculation
indicates that it does not yet have adequate convergence \cite{Andersen3}.
As another approach,
3D reduction theory has also been widely studied and 
has yielded some promising arguments \cite{Kajantie}, 
but this method, which is defined 
only for the high temperature limit, 
cannot be applied to confinement/deconfinement physics.

For reliable phenomenological analyses of 
high energy heavy-ion collisions,
it is important to obtain information on the magnetic and electric
masses of gluons nonperturbatively; 
a numerical study of lattice QCD as a first-principles calculation
should play an important role here. 

There have been many lattice studies of finite-temperature QCD,
but only a few calculations of the electric and magnetic screening 
masses can be found in the literature \cite{Ogilvie}, 
other than for the case of color $SU(2)$ \cite{Heller,Cucchieri}.
The electric mass has been estimated from the Polyakov loop correlation
 functions to obtain the screened heavy-quark-antiquark potential
 \cite{Irback,Gao,Kaczmarek}.

The main aim of our study is to 
obtain reliable $SU(3)$ electric and magnetic masses through
large-scale lattice QCD simulations
and to reveal their temperature dependence \cite{saito}.
We also compare our numerical data with 
the predictions of leading order perturbation (LOP), 
HTL resummation, and other analyses.

Since our mass extraction is based on measuring a gauge dependent 
gluon propagator, it is important to check the gauge invariance of
both screening masses \cite{Cucchieri2}. 
This test is essential particularly for the magnetic mass 
because it has a poor definition in the frame of the perturbation.

Gluons are essential ingredients in QCD dynamics, and QCD undergoes
a phase transition from the confinement to the deconfinement phase
when the temperature increases.
Therefore, we expect gluon propagators to show different behavior in each
phase, and their study provides information on
 confinement/deconfinement dynamics \cite{Nakamura2}.

We measure gluon propagators, which depend on the gauge used, 
and therefore a gauge-fixing procedure is indispensable.
However, gauge fixing on the lattice is difficult practically 
and conceptually.
Usually, gauge fixing is carried out by the iterative technique \cite{Mandula},
and it is very time consuming.
The conceptual difficulty is that the gauge is not uniquely fixed;
this is known as the Gribov copy problem \cite{Gribov}.
In order to overcome these difficulties, we adopt here
a stochastic quantization with Zwanziger's gauge-fixing term
 \cite{Zwanziger,Seiler,Mizutani},
instead of the path integral method.      

In this paper, 
we extend our previous results
from the $20^2\times 32 \times 6$ lattice simulations \cite{saito}
to the level that the present computer power can reach;
we add more detailed values for the electric and magnetic masses
and results at higher temperature 
from the larger lattice $32^2\times 48 \times 6$ simulation. 
We also give a detailed description of the algorithm employed
in this study.
In Sec. II, we describe the stochastic gauge quantization 
together with the Gribov copy problem.
The definitions of the gluon propagator 
and electric and magnetic masses are given.
A large part of Sec. III is devoted to our simulation results on 
the small lattice $20^2\times 32 \times 6$.
First we describe all input parameters of the simulation
and the statistics needed to measure reliable gluon propagators.
Then we show the gluon behavior and 
the electric and magnetic masses extracted from it.
The gauge dependence check and 
temperature dependence for both screening masses are also given.
Finally, we compare the numerical data with the perturbative argument,  
 add the higher temperature result, and comment on 
the finite volume effect.
Section IV gives the conclusions.
 
The main part of the calculation was carried out 
on the SX-5(NEC) vector-parallel computer of RCNP(CMC), Osaka University.
We used a parallel queue with 4, 8 and 16 CPUs and
required about six months to complete this work. 

\section{Stochastic gauge fixing on lattices
and gluon propagators}
\

\subsection{Lattice gauge action}

The lattice regularization scheme of QCD 
is the gauge invariant Euclidean theory which enables us to perform
a nonperturbative calculation based on the Monte Carlo
numerical technique \cite{Wilson}.
The simplest standard Wilson gauge action of lattice QCD 
can be defined from the continuum QCD as 
\begin{equation} 
S_g=\beta \sum ( 1 - \frac{1}{3} \mbox{Re Tr}
\left[
U_{\nu}(x)U_{\mu}(x+\hat{\nu})
U^{\dagger}_{\nu}(x+\hat{\mu})U^{\dagger}_{\mu}(x) \right] ),\\
\hspace{0.2cm} \beta=\frac{2N_c}{g^2}.\label{action}
\end{equation} 
Here a link variable, $U_\mu(x)$, stands for the $SU(3)$ color gauge field:
\begin{equation}
U_{\mu}(x)= e^{iga_{\mu} A_{\mu}(x) },
\end{equation}
where $a_{\mu}$ is the lattice spacing, i.e, the lattice cutoff, 
and $A_{\mu}$ represents the gauge potential of the gluon.
In this study, we adopt the quenched lattice simulation 
(pure gauge QCD) without a dynamical quark effect, using Eq. (\ref{action}).

\subsection{Gauge fixing and Gribov copy on lattices}

We are interested in a direct calculation of the gluon propagator, 
and the extraction of electric and magnetic screening masses from it.
Therefore we must fix the gauge of the gluon fields on the lattice, 
where the gauge transformation is given by
\begin{equation}
U_{\mu}(x) \rightarrow \omega^{\dagger}(x)
           U_{\mu}(x) \omega(x+\hat{\mu}).
\end{equation}
$\omega$ stands for a gauge rotation matrix $\in SU(3)$ on the lattice.

In this study, we focus on a Lorentz-type gauge, which is defined in the
continuum as 
\begin{equation}
\partial_{\mu} A_{\mu}(x) = 0,
\end{equation}
while in the discrete lattice theory
\begin{equation}
\Delta^a(x) \equiv \sum_{\mu=1}^4 2 \mbox{ Im Tr }t^a 
\{U_{\mu}(x) - U_{\mu}(x-\hat{\mu})\} = 0.
\label{lgfc}
\end{equation}
Here $t^a$ is the $SU(N)$ generator with the relation 
$\mbox{Tr} [t^a t^b] = \frac{1}{2} \delta^{ab} $.
The above condition (\ref{lgfc}) is equivalent to
\begin{equation}
\begin{array}{lll}
\delta_{\omega} I &=& 0,\\
I&\equiv& \sum_{x,\mu} \mbox{Re Tr} \omega^{\dagger}(x) U_{\mu}(x)
\omega(x+\hat{\mu}).\label{cond1}
\end{array}
\end{equation}
Wilson \cite{Wilson80} and Mandula and Ogilvie \cite{Mandula} suggested
the following condition
for the gauge fixing method on the lattice:
\begin{equation}
\mbox{Max.}_{\omega} I.\label{cond2}
\end{equation}
The continuum version of Eq. (\ref{cond2}) was discussed in Refs.
 \cite{Maskawa} and \cite{Semenov}.
\begin{figure}[hbt]
\begin{center}
\scalebox{0.6}{
\includegraphics{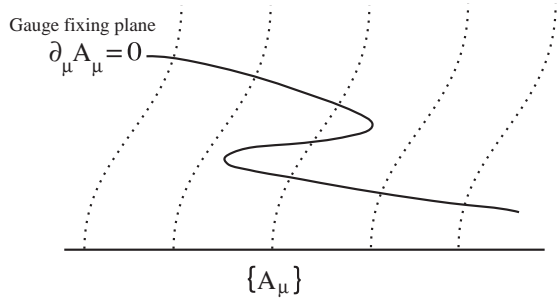} }
\caption{Gribov \cite{Gribov} pointed out that   
a gauge is not uniquely fixed for non-Abelian theories,
 which is called the Gribov copy problem. 
This does not appear in the  perturbative frame.
The existence of the Gribov copy has been confirmed
in several works \cite{Nakamura,Hioki,Bornyakov}.
\label{gribov-pic}}
\end{center}
\end{figure}

The conditions (\ref{cond1}) and (\ref{cond2})
are not equivalent to each other since there may be
 local maxima or minima of $I$ which satisfy  Eq. (\ref{cond1}).
The gauge-fixing configuration cannot be uniquely fixed; 
this is called the ``Gribov copy'' \cite{Gribov} 
and is illustrated in Fig. \ref{gribov-pic}.
If we study this problem using a numerical lattice simulation
based on the iterative procedure,
it is very difficult to find a true maximum.

\subsection{Stochastic gauge fixing}

\begin{figure}[hbt]
\begin{center}
\scalebox{0.60}
{ \includegraphics{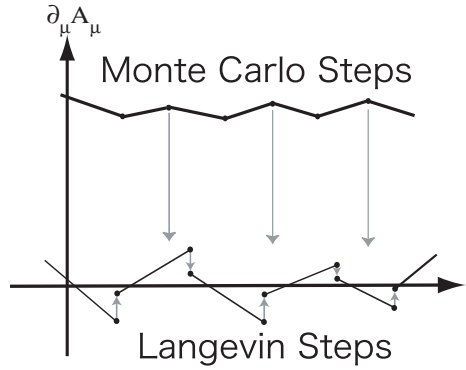} }
\caption{
The figure illustrates the gauge-fixing procedure by 
the Wilson-Mandula-Ogilvie iterative method and by stochastic gauge fixing.
Gray arrows indicate the gauge rotation for each algorithm.
In the first algorithm, 
Monte Carlo update steps are performed without restriction
on the $\partial_{\mu} A_{\mu}=0 $ plane.
When a gauge fixed configuration is needed, the gauge configuration 
is rotated to the gauge-fixed plane, $\partial_{\mu} A_{\mu} =0 $, 
by iteration \cite{Mandula}.
On the other hand, in the case of Langevin gauge fixing,
 configurations are updated by fluctuation 
around $\partial_{\mu} A_{\mu} = 0 $.
\label{gf}}
\end{center}
\end{figure}


Our approach to the gauge-fixing procedure 
is to use the stochastic gauge quantization
instead of the Monte Carlo path integral.
The stochastic quantization is 
based on the Langevin equation which introduces virtual time 
in addition to the Euclidean coordinate.
Zwanziger introduced a gauge-fixing term as
\begin{equation}
\frac{dA_{\mu}^a}{d\tau}
= - \frac{\delta S}{\delta A_{\mu}^a}
+\frac{1}{\alpha} D^{ab}_{\mu}(A)
\partial_{\nu} A_{\nu}^b
+\eta_{\mu}^a,\label{sgqe}
\end{equation}
where $D_{\mu}^{ab}(A)$ is
a covariant derivative, $\tau$ stands for Langevin time,
and $\eta$ is a Gaussian noise term.
The second term on the right-hand side is a gauge-fixing term.
$\alpha$ is a gauge parameter; $\alpha=0$ corresponds to
the Lorentz gauge and $\alpha=1$ to the Feynman gauge.

Mizutani and Nakamura \cite{Mizutani} developed the
lattice version of the stochastic gauge fixing.
The link variables are rotated through the following
gauge transformation depending on the virtual time: 
\begin{equation}
U_{\mu}(x,\tau+\Delta\tau) =
\omega^{\dagger}(x,\tau)e^{if_{\mu}^a t^a}
U_{\mu}(x,\tau) \omega(x+\hat{\mu},\tau).
\label{LatSQeq}
\end{equation}
Here $f$ stands for the force   
\begin{equation}
f_{\mu}^a=-\frac{\partial S}{\partial A_{\mu}^a} \Delta \tau
+ \eta^a \sqrt{\Delta \tau},
\end{equation}
and the gauge rotation matrix is given by
\begin{equation}
\omega = e^{i\beta \Delta^a t^a \Delta \tau /\alpha }.
\label{g-r}
\end{equation}
If $\omega = I$, Eq. (\ref{LatSQeq}) is a lattice Langevin process. 
Gauge rotation, Eq. (\ref{LatSQeq}), with Eq. (\ref{g-r}), leads
to the gauge fixing term as $\Delta \tau \rightarrow 0$.
Equation (\ref{LatSQeq}) means that
 the gauge rotation and Langevin step are executed alternately, 
as illustrated in Fig. \ref{gf}.

\begin{figure}[hbt]
\begin{center}
\scalebox{0.3}{
\includegraphics{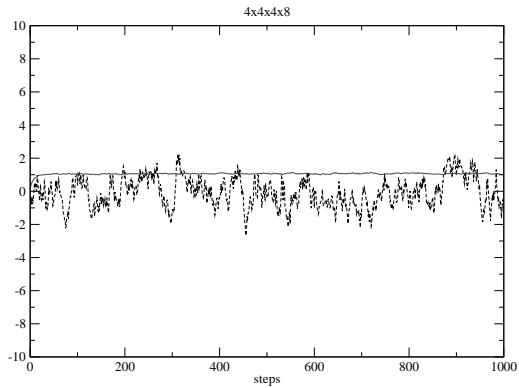} }
\caption{These data are produced on the lattice $4^3 \times 8$
with $\Delta \tau = 0.01$(Langevin step interval), $\alpha = 1.0$
(gauge parameter) at $\beta=6.0$.
The solid line means the average of 
$\sum_{x,a} \{ \Delta^a(x) \}^2$, while
the dashed line is the value of $\Delta^1(1)$. 
$\Delta^a$ fluctuates around $\Delta^a=0$ and gauge configurations
are sufficiently fixed on the $\partial_{\mu} A_{\mu}=0$ plane.  
}\label{dudmu}
\end{center}
\end{figure}

In this stochastic quantization with Lorentz-type gauge fixing,
$\Delta^a$ fluctuates around the gauge-fixing plane $\Delta^a=0$.
For example, when we take $\alpha=1.0$ and $\Delta \tau = 0.01$,
 $\Delta^a$ on $4^3\times 8$ with $\beta=6.0$
behaves as shown in Fig. \ref{dudmu}.
We confirm that $\Delta^a$ fluctuates around $\Delta^a=0$
which indicates that good gauge fixing is achieved. 

There are two reasons for using the stochastic gauge-fixing method in
this study. One is a practical issue. 
When we use the standard Wilson-Mandula gauge-fixing method \cite{Mandula},
 the iterative procedure is applied to accepted gauge configurations 
for each Monte Carlo step in Fig. \ref{gf}.
Then the number of iterations is unpredictable, particularly 
for large lattices.
On the other hand, in algorithm (\ref{LatSQeq}),
we simultaneously repeat the step of update and gauge rotation
in Fig. \ref{gf}
and are free from the convergence problem of gauge fixing.
Therefore we can estimate the CPU time precisely.
Moreover, this algorithm, in which the gauge parameter $\alpha$ 
can be changed at will, 
is advantageous when testing gauge invariance.

A conceptual problem is the Gribov ambiguity \cite{Gribov}.
The algorithm has a noteworthy feature, i.e., the second term 
of Eq. (\ref{sgqe}) gives rise to a configuration such that
\begin{equation}
\begin{array}{lll}
\frac{d}{d\tau}\sum_x\Delta^2&<&0\mbox{\ in\ \ \ \ \ \ } \Omega \\
                             &>&0\mbox{\ out of\ } \Omega.
\end{array}
\end{equation}
Here $\Omega$ stands for the Gribov region,
\begin{equation}
\mbox{Gribov\ region}: \Omega \equiv
 \{ A_{\mu}^a(x) | \partial_{\mu} A_{\mu}^a(x)=0,
 - \partial_{\mu} D_{\mu}^{ab} > 0 \}.
\end{equation}
That is, the stochastic gauge-fixing term is attractive (repulsive)
inside (outside) the Gribov region \cite{Zwanziger,Seiler}
if we start from the trivial configuration $\{A_{\mu}=0\}$.
Although our algorithm may not completely eliminate copies,
we conclude that it is a more effective method.

\begin{figure}[hbt]
\begin{center}
\scalebox{0.3}{
\includegraphics{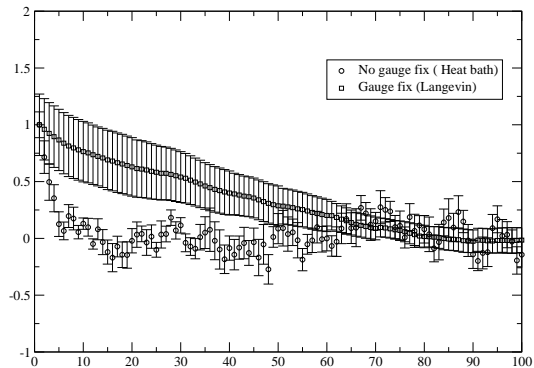} }
\end{center}
\caption{
Autocorrelation of the Polyakov loops for the Langevin algorithm
with the stochastic gauge fixing ($\Delta \tau = 0.03$)
 and the pseudo-heat-bath update algorithms 
as a function of the iteration steps.
This calculation is done at $\beta=6.1$ on the $8^3\times 4$ lattice.
}
\label{PolAuto}
\end{figure}

Since the update algorithm described here is not as popular as the Metropolis
or pseudo-heat-bath method, we show the autocorrelation of the Polyakov lines 
$L(\tau)L(\tau+n\tau)$ in Fig. \ref{PolAuto} together with that of 
the pseudo-heat-bath.  

\subsection{Definition of gluon propagators}

We define the gauge field $A_{\mu}(x)$
in terms of the link variables as 
\begin{equation}
A_{\mu}(x) = 2 \sum_a t^a \mbox{Im} \mbox{Tr} t^a U_{\mu}(x).
\end{equation}
We calculate massless gluon correlation functions with finite momentum,
\begin{equation}
G_{\mu\nu}(p_x,p_y,p_t,z)=
\langle \mbox{Tr}
A_\mu(p_x,p_y,p_t,z)A_\nu(-p_x,-p_y,-p_t,0) \rangle\label{correl}.
\end{equation}

In our study, to avoid a mixture of the longitudinal and transverse modes,
we adopt the transverse conditions $p_i\cdot A_i = 0$
and measure the partially Fourier transformed propagator including
one momentum $p_x=2\pi/N_x$ or 
$p_y=2\pi/N_y$ \cite{Nakamura}.
We will obtain the gluon mass using a lattice energy momentum relation.
This is different from other calculations \cite{Heller,Ogilvie}, 
where the mass extraction from zero momentum propagator
was studied.

In thermal perturbative QCD \cite{Gross,Kapusta,Bellac}, 
the electric mass is defined from the temporal part
of the gluon polarization tensor $\Pi_{\mu\nu}$,
while its spatial part is considered as the magnetic mass. 
Thus we can construct an electric propagator using 
a temporal one:

\begin{equation}
G_e(p,z) \sim
\frac{1}{2}
\left[
             G_{tt}(\frac{2\pi}{N_x},0,0,z)
            +G_{tt}(0,\frac{2\pi}{N_y},0,z)
\right].\label{eprop}
\end{equation}
In the same way, a magnetic propagator is defined by spatial components:
\begin{equation}
G_m(p,z) \sim
\frac{1}{2}
\left[
             G_{xx}(0,\frac{2\pi}{N_y},0,z)
            +G_{yy}(\frac{2\pi}{N_x},0,0,z)
\right].\label{mprop}
\end{equation}

Because the screening mass is given as a pole of
the denominator of the momentum space propagator, 
$1/(p^2+m^2)$,
$G_e(z)$ and $G_m(z)$ are expected to behave as 
exponential damping functions in the z direction 
on distances, $Z \geqslant 1/T$ with the masses,
\begin{equation}
   G_{e(m)}(z) \sim e^{-E_{e(m)}(p)z}.\label{exp}
\end{equation}

\section{Results}

\subsection{Simulation parameters}

We use mainly the lattice of the size $N_xN_yN_zN_t=20^2\times 32 \times 6$
which satisfies the condition $N_x \geqslant 3N_t$. 
On this lattice the long range area corresponds to 
$z \geqslant 1/T$ and thus reliable information for screening physics
may be produced. 

We summarize lattice cutoff values and the corresponding temperature
in Table \ref{para}.
Varying $\beta$, i.e., the lattice coupling constant,
we change the temperature $T=1/N_ta$.
 The pure gauge lattice with $N_t=6$  
has critical $\beta_c \sim 5.89$ and we adopt 
 $T_c \sim 256$ \cite{Boyd} as the critical temperature.
To estimate the lattice cutoff value, we employ the results in
 Refs. \cite{QCD_TARO,Allton}.

In addition,
we prepare a lattice of the size $32^2 \times 48 \times 6$
in order to investigate the finite size effect
of screening masses, and to obtain them at higher temperature.

\begin{table}[t]
\caption{ We estimate a lattice cutoff and its temperature scale
 by using mainly QCD\_TARO fit function \cite{QCD_TARO,Allton}.
We adopt $T_c \sim 256$ \cite{Boyd}.}\label{para}
\vspace{0.25cm}
\begin{center}
\renewcommand{\arraystretch}{1.00}
\begin{tabular}{lcccccccc}
\hline
\hline
$\beta$ &$a^{-1}$[GeV] & T[MeV] & $T/T_c$&
$\beta$ &$a^{-1}$[GeV] & T[MeV] & $T/T_c$\\
\hline
5.8 &1.33&222 &0.86 &6.4 &3.52&586 &2.29\\
5.90&1.62&270 &1.05 &6.5 &4.12&690 &2.69\\
5.95&1.77&295 &1.15 &6.6 &4.60&767 &2.99\\
6.0 &2.04&340 &1.32 &6.7 &5.24&874 &3.41\\
6.05&2.09&349 &1.36 &6.8 &5.96&994 &3.88\\
6.1 &2.27&378 &1.47 &6.9 &6.76&1128&4.40\\
6.2 &2.64&447 &1.74 &7.0 &7.64&1274&4.97\\
6.3 &3.05&509 &1.99 &7.1 &8.61&1436&5.61\\
\hline
\hline
\end{tabular}
\end{center}
\end{table}
\vspace{1cm}

\subsection{Necessary statistics to obtain reliable gluon propagators}
\

\begin{figure}[hbt]
\begin{center}
\scalebox{0.300}{
\includegraphics{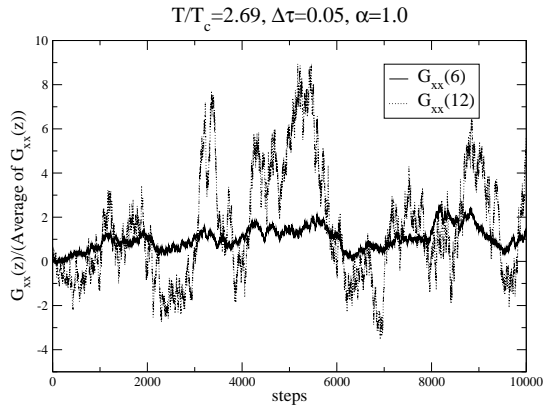}%
}
\caption{
Typical gluon propagator behavior as a function of Langevin steps
on the lattice $ 20^2 \times 32 \times 6 $.
We find that fluctuations of $G_{xx}(12)$ are much larger than
those of $G_{xx}(6)$.
In order to investigate the screening effect,
the long range contribution should be adopted.
Consequently, we need a large number of statistics.
\label{hist3}}
\end{center}
\end{figure}


We observe a large fluctuation of the gauge propagators,
 particularly at long distances, i.e.,
we suffer from a long autocorrelation time.  
We show in Fig. \ref{hist3} the typical behavior
of gluon propagators $G(z)$ as a function of the Langevin step.
In order to analyze the gluon propagator and 
measure the screening masses we require $0.2-0.4\times10^6$ steps
as the typical number of simulation data.


\subsection{Gluon propagators}

Although the gluon propagator itself is gauge dependent,
it gives us some insights into
the gluon dynamics of the confinement/deconfinement physics.

\begin{figure}[hbt]
\begin{center}
\scalebox{0.300}{
\includegraphics{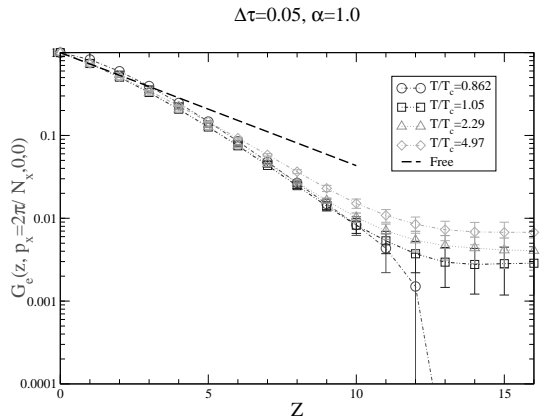} }
\caption{The electric propagator in the confinement (circle) and
deconfinement regions (other symbols)
with $P_{x(y)}=2\pi/N_{x(y)}$.
All propagators are found to become massive
compared with the free propagator (long dashed line).
In the confinement regions,
the propagator at long distances behaves like a very massive
particle and vanishes, whereas the propagator beyond $T_c$ has a finite mass.
\label{prop-e}}
\end{center}
\end{figure}

The electric propagator is shown in Fig. \ref{prop-e}
\footnote{In the following,
all the propagators are normalized at $z=0$,
namely they are divided by $G_{e(m)}(z=0)$ 
to compare each other.},
where the free massless propagator is also shown by the dashed line. 
Gluons at short distances, i.e., $z<6=1/T$,
have very similar behavior to the free propagator.
However, at long distances, gluon propagators
decrease more rapidly than the free one.
This indicates that the electric screening mass does not vanish 
at all temperatures.

Comparing data below and above $T_c$, we 
find that the dynamics of the gluon is completely different
in the confinement and deconfinement regions.
The electric gluon mass in the confinement region
becomes heavier;
the electric gluon is completely screened in the confinement regions.
This was first observed in Ref. \cite{Nakamura2}.
Consequently, we cannot employ the assumption Eq. (\ref{exp}).
On the contrary, the propagator in the deconfinement regions
decreases exponentially even at long distances with
a finite mass.

\begin{figure}[hbt]
\begin{center}
\scalebox{0.300}{
\includegraphics{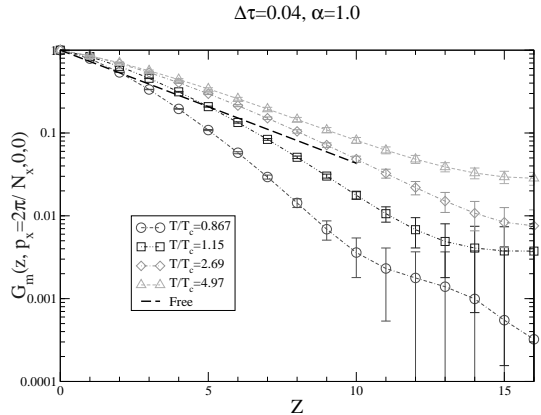} }
\caption{
The magnetic gluon propagator in the confinement (circle) and
deconfinement regions (other symbols) with $P_{x(y)}=2\pi/N_{x(y)}$.
Although the magnetic gluon propagators in this figure 
have large errors at long distances, they seem to have 
similar behavior to the electric part, except for the short distance
 behavior in the deconfinement regions. 
\label{prop-m}}
\end{center}
\end{figure}

Similar behavior is seen in the magnetic parts in Fig. \ref{prop-m};
although the long distance magnetic gluons have large errors here,
all magnetic gluons are found to be massive at long distances  
in the confinement/deconfinement phase, and
 their effective mass in the confinement phase is heavier. 
We notice that the short distance behavior of magnetic gluons
in the deconfinement regions looks unconventional.
The magnetic gluon propagators follow a convex curve at short distances 
while the electric ones do not.
The magnetic gluons
behave as if they had an imaginary screening mass or 
negative spectral function, which may be the reason why
the magnetic mass is not screened
at least in LOP calculation.

As clearly seen in Fig. \ref{prop-m} and as we discuss below,
the effective mass of the magnetic mass is $z$ dependent.
We take  the value around $z = 1/T$ as the electric case, since
it is a relevant quantity at finite temperature screening.

\begin{figure}[hbt]
\begin{center}
\scalebox{0.400}{
\includegraphics{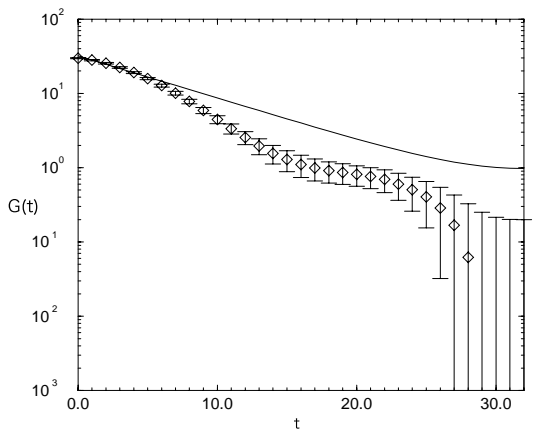} }
\caption{
Transverse gluon propagator, $G_T(t)$ with 
${\bf p}=((2\pi/N_x),0,0)$, on $48^3 \times 64$ 
at $\beta=6.8$ (confinement phase). 
Solid line represents the free propagator \cite{Nakamura2}.
\label{zero} }
\end{center}
\end{figure}

Gluons are essential ingredient of QCD but they are confined below $T_c$.
As shown in Fig. \ref{zero},
 their propagator is convex upward at several regions.
This is possible only when the spectral function is {\it not} positive definite.
This peculiar behavior was first observed in Ref. \cite{Mandula} and confirmed
in Ref. \cite{Nakamura95}.  
The feature does not contradict the fundamental postulate
of quantum field theories, because gluons in the confinement region are
not physically observable particles.  
Instead, this is a glimpse of the confinement mechanism in the infrared
region which is still far from our understanding.
Gribov's conjecture for the gluon propagator 
\begin{equation}
G(p) \sim \frac{Z}{
p^2 + b^4/p^2 },
\end{equation}
vanishes at $p^2=0$, and its Fourier transformation to the coordinate space 
is not convex downward,
\footnote{We thank E. Seiler and D. Zwanziger for helpful discussions on this point.}
\begin{equation}
\tilde{G}(t) \sim \frac{Z}{r} e^{rt\cos\phi} \cos(rt\sin\phi+\phi),
\end{equation}
where
$r \equiv (|\vec{p}|^4+b^4)^{1/4}$
and
$\phi\equiv\frac{1}{2}\tan^{-1}({b^2}/{\vec{p}^2})$.
The upward convex shape of the gluon propagators in the deconfinement
region may provide us with hints about the glue dynamics.

Below the critical temperature $T_c$, we obtain a similar result 
even for electric gluons at short distances.
This seems natural since the perturbative argument 
in the confinement regions is generally not suitable
and the confining correlation function would also give 
the negative spectral function.

\subsection{Mass as a pole}

To obtain the screening mass from the propagators,
the following formula is used:
\begin{equation}
 G_{e(m)} \sim \cosh[E_{e(m)}(p)(z-N_z/2)].
\end{equation}
We employ data for $z \geqslant 1/T(=N_ta)$,
because the screening effect occurs at sufficiently long distances. 
All fittings are done from $z=6$ to $N_z/2$, 
and $\chi^2/NDF \sim O(1)$,  where $NDF$ indicates the number of degrees
 of freedom.

\begin{figure}[hbt]
\begin{center}
\scalebox{0.40}{
\includegraphics{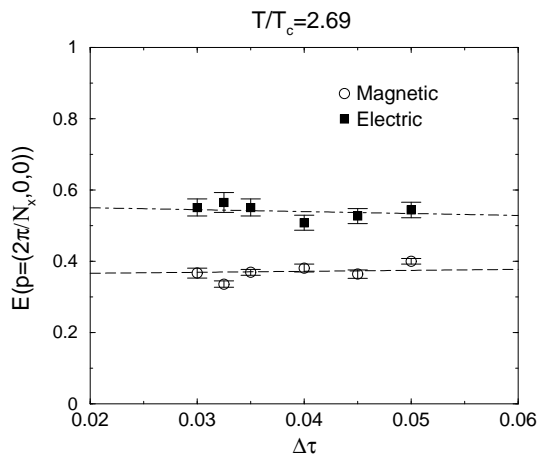} }
\caption{$\Delta \tau$ dependence of masses is slight.
Finally to obtain a final value at $\Delta \tau=0$,
we use the linear function for the extrapolation.
\label{dtdeps}}
\end{center}
\end{figure}

To obtain the final result at $\Delta \tau=0$,
we must extrapolate the data with respect to Langevin step width.
The Runge-Kutta algorithm is applied to 
reduce the finite Langevin step $\Delta \tau$ dependence \cite{Batrouni}.
We perform simulations for a set of parameters with
$\Delta\tau = 0.03 \sim 0.05$.
Table \ref{deltatau} and Fig. \ref{dtdeps} represent
$E(p)$ measured here versus $\Delta \tau$.
The slight dependence of $\Delta \tau$ enables us
to use a linear function when fitting data.
We finally obtain the mass from $E(p)$ by the following
lattice energy-momentum relation \cite{Heller}
\footnote{We ``assume" this relation to extract the mass.}:
\begin{equation}
\sinh^2\frac{Ea}{2}=\sinh^2\frac{ma}{2}
+\sum_{i=1}^{3}\sin^2\frac{p_i a}{2}.
\end{equation}
\begin{table}[htb]
\caption{
This shows the typical example of the Langevin step dependence
at $T/T_c=2.69$.
For all simulations for $\Delta \tau = 0.03 - 0.05 $,
approximately $0.2-0.4\times 10^6$ steps (measurements) 
are used after eliminating about $3000-5000$ steps 
as thermalization.
``${\bf p}\neq 0$'' means including 
the momentum $P_{x(y)} = 2\pi/N_{x(y)}$.
We extrapolate these data to $\Delta \tau = 0$ and
 then obtain $m_e a (p_\mu=0)=0.470(38)$ and
 $m_m a (p_\mu=0)=0.199(52)$ using lattice energy-momentum relation.
}
\vspace{0.25cm}
\begin{center}
\renewcommand{\arraystretch}{1.00}
\begin{tabular}{lcccc}
\hline
\hline
$\Delta \tau$&Number of steps  
& $m_m a({\bf p}\neq 0) $
& $m_e a({\bf p}\neq 0) $\\
\hline
0.05  &275000&0.400(08)&0.544(22)\\
0.045 &240000&0.364(12)&0.527(21)\\
0.04  &380000&0.381(11)&0.508(21)\\
0.035 &280000&0.369(08)&0.551(24)\\
0.0325&300000&0.336(09)&0.565(28)\\
0.03  &320000&0.367(14)&0.551(24)\\
\hline
0.00  &   &0.369(46)&0.561(40)\\
\hline
\hline
\end{tabular}
\end{center}\label{deltatau}
\end{table}

\subsection{Gauge invariance}

\begin{figure}[hbt]
\begin{center}
\scalebox{0.40}{
\includegraphics{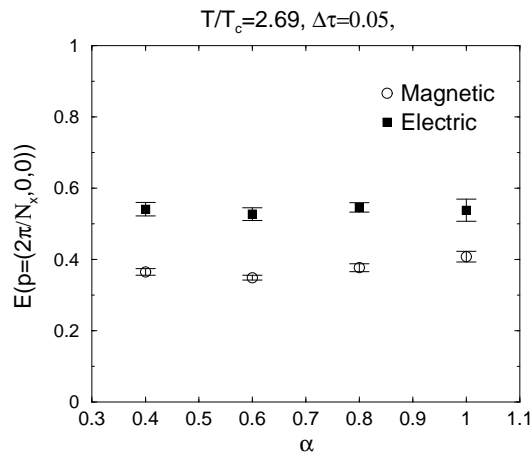}%
}
\caption{Gauge dependence for electric and magnetic
screening masses. Gauge dependence
of both screening masses is very slight.
\label{meg}}
\end{center}
\end{figure}

The screening mass is physical and expected to be gauge invariant.
However, since the gluon propagators defined by
Eqs. (\ref{eprop}) and (\ref{mprop}) are gauge dependent,
it is important to check
whether the screening masses obtained here
are gauge invariant or not.
In addition, since the magnetic mass cannot be defined
by a perturbative calculation, 
it is particularly important to check its gauge dependence.
In Fig. \ref{meg}, we show the gauge parameter $\alpha$ dependence of
electric and magnetic masses.
The gauge dependence of both screening masses is 
found to be very slight, namely, the result strongly suggests
 that they are gauge invariant and physical observables.

\subsection{Temperature dependence}

\begin{figure}[hbt]
\begin{center}
\scalebox{0.30}{ \includegraphics{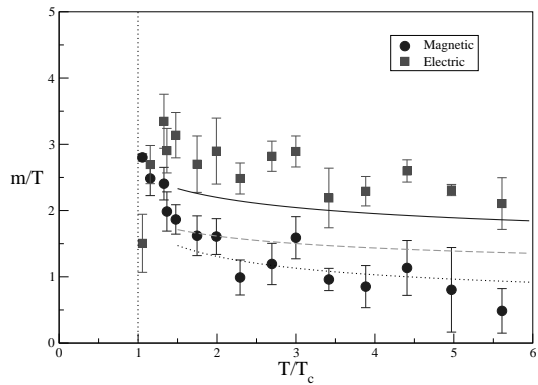} }
\caption{Temperature dependence of electric and magnetic screening masses.
The dotted line is fitted by the assumption, $m_g \sim g^2T$.
For the electric mass,
the dashed and solid lines represent LOP and HTL resummation results,
respectively.
\label{emt}}
\end{center}
\end{figure}

We study the temperature dependence of the screening mass
in the range $T/T_c=1-6$ which would be realized
in high energy heavy-ion collision experiments 
such as RHIC or LHC \cite{Muller}.
Table \ref{tempdeps} and Fig. \ref{emt} 
show electric and magnetic masses as a function of the temperature.
The magnetic part definitely has nonzero mass in this
temperature region.
As $T$ increases, both $m_{e(m)}/T$ decrease monotonically,
and at almost all temperatures, the magnetic mass is less 
than the electric one,
except very near $T_c$ where the electric mass decreases very quickly
as $T$ approaches $T_c$.

\begin{table}[t]
\caption{Temperature dependence of the
electric and magnetic masses 
which are extrapolated to the Langevin step $\Delta \tau=0$.
 \label{tempdeps}}
\vspace{0.25cm}
\renewcommand{\arraystretch}{1.00}
\begin{center}
\begin{tabular}{cccccc}
\hline
\hline
$T/T_c$ & $m_e/T$ & $m_m/T$ & $T/T_c$ & $m_e/T$ & $m_m/T$ \\
\hline
1.05 &1.506(438) &2.802(054) &2.69 &2.820(228) &1.194(312) \\
1.15 &2.694(288) &2.484(258) &2.99 &2.892(234) &1.590(318) \\
1.32 &3.348(408) &2.406(246) &3.41 &2.190(450) &0.960(168) \\
1.36 &2.904(336) &1.986(296) &3.88 &2.292(222) &0.852(318) \\
1.47 &3.138(342) &1.866(222) &4.40 &2.598(168) &1.134(414) \\
1.74 &2.700(426) &1.620(300) &4.97 &2.310(084) &0.804(638) \\
1.99 &2.898(498) &1.608(270) &5.61 &2.106(390) &0.486(336) \\
2.29 &2.484(234) &0.990(264) &     &           &           \\
\hline
\hline
\end{tabular}
\end{center}
\vspace{0.5cm}
\end{table}

\subsection{Comparison with LOP and HTL resummation results}

We perform a fitting analysis for our numerical results 
using the following ansatz:
\begin{equation}
\displaystyle
 \frac{m_e}{T} = C_e g(T),\hspace{0.25cm}
 \frac{m_m}{T} = C_m g^2(T),\label{scale}
\end{equation}
whose g dependence is predicted by the perturbative and
3D reduction analysis \cite{Kapusta,Bellac}
and we assume $C_e$ and $C_m$ are free parameters. 
In the following discussion, 
the data above $T \sim 1.5T_c$ are used.
Here we use the running couplings as
\begin{equation}
g^2(\mu)=\frac{1}{2b_0\log(\mu/\Lambda)}
\left(1-\frac{b_1}{2b_0}\frac{\log(2\log\mu/\Lambda)}
{\log(\mu/\Lambda)} \right),
\end{equation}
and we set $\mu=2\pi T$, which is the Matsubara frequency as the
renormalization point and $\Lambda = 1.03T_c$ \cite{Andersen}
as the QCD mass scale.
$b_0$ and $b_1$ are the first
two universal coefficients of the renormalization group,
\begin{equation}
b_0=11N_c/48\pi^2,\quad b_1=(34/3)(N_c/(16\pi^2))^2.
\end{equation}
As a result, we obtain 
\begin{equation}
\begin{array}{clc}
C_e&=1.63(3), &\chi^2/\mbox{NDF}=0.715,\\
C_m&=0.482(31),& \chi^2/\mbox{NDF}=0.979.\label{fitdata1}
\end{array}
\end{equation}
The scalings expected in Eq. (\ref{scale}) for electric and
magnetic masses are found to work well.
However, the magnitude of $C_e$ is larger
than $C_e^{LOP}=1$. 
On the other hand, for magnetic mass, 
a self-consistent inclusion technique in Ref. \cite{Alexanian}
gives $m_g=0.568 g^2 T$, which is close to our fitting result.

The HTL resummation technique
applying the free energy of hot gluon plasma
has been widely discussed \cite{Andersen,Andersen2}. 
Rebhan gave a formula for the electric mass in the one-loop
HTL perturbation theory \cite{Rebhan} and
for the case of $SU(3)$, 
\begin{equation}
\displaystyle
m_e^2=m_{e,0}^2
\left[ 1 + \frac{3g}{2\pi}\frac{m_e}{m_{e,0}}
\left( \log\frac{2m_e}{m_m} - \frac{1}{2} \right)+O(g^2)
 \right], m_{e,0}=gT.
\end{equation}
Here we assume the magnetic mass to be of the order of $g^2$.
Substituting our fitted value for $m_m$,
we can solve the above equation iteratively.
In Fig. \ref{emt},
 we show this HTL resummation together with the LOP result.
The HTL result gives a better description
 than the naive perturbation,
upon comparing with our numerical experiment.

The electric mass was obtained also by
a heavy $q\bar{q}$ potential from the $SU(3)$ Polyakov loop correlator
at finite temperature in Refs. \cite{Gao,Kaczmarek}. 
Our results here are inconsistent with 
theirs, since the mass extraction from the heavy $q\bar{q}$ potential
 cannot be consistently performed due to ambiguity of its 
fitting assumption.
In addition, a 3D reduction argument \cite{Kajantie}
has shown that $m_e/gT$ goes down when $T$ increases, but
even at $T\sim 1000\Lambda_{\overline{MS}}$
the electric mass is still about $3m_{e,0}$.
This observation agrees qualitatively with our analysis.

\subsection{Higher T and finite size effect\label{mht}}

\begin{figure}[hbt]
\begin{center}
\scalebox{0.30}{
\includegraphics{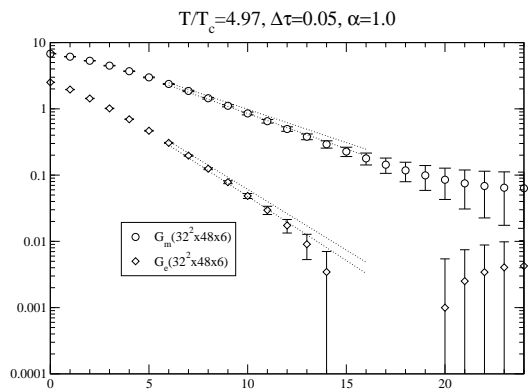} }
\caption{Typical electric and magnetic propagators 
on the large lattice $32^2\times48\times6$ (open),
and the data for $20^2\times32\times6$ (band between dotted lines with
error) are reproduced by using the values in Table \ref{tempdeps}.
Both calculations give very similar value
for electric and magnetic parts. \label{lprop} }
\end{center}
\end{figure}

Although the main result in this paper is based on 
studies for the small lattice size $20^2 \times 32 \times 6$
as discussed in the previous section, 
we additionally perform the simulation 
on the large lattice $32^2 \times 48 \times 6$
to go to higher temperature regions and 
to check the finite-size effect of
the screening masses. However, as the lattice size increases,
the behavior of the long distance gluon 
cannot be controlled because of a large fluctuation.
A typical result on the large lattice 
is shown in Fig. \ref{lprop}.
 Even after the $0.3-0.4\times 10^6$ measurements,
we could not determine the electric gluon propagator at
long distances ($ z \geqslant 16 $), while the magnetic gluon is 
properly correlated.\footnote{
Magnetic propagator determination is also difficult
near $T_c$.}
Nevertheless, provided that we adopt only 
the data for the intermediate regions above $z=6$ 
until the disappearance of the propagator,
we obtain similar results for the electric and magnetic masses
as seen in Fig. \ref{lprop} in the same temperature regions.

Using the criterion describing above, we may consistently obtain
 both screening masses on the large lattice,
 and can argue that the magnetic mass has 
a stronger finite-size effect than the electric one.
$E(p)$ appearing in Eq.(17) are shown in Table \ref{hight}
at $T/T_c=8.99$ and $16.12$.
The momenta for the small and large lattices
are $p_x=2\pi/N_x\sim 0.314$ and $p_x=2\pi/N_x\sim 0.196$, respectively; 
namely, the result on the small lattice
implies that the magnetic mass at high temperature ($T/T_c>5$)
seems to be going to zero, while on the large lattice it remains finite.
\footnote{Note that the data of $\beta=7.5$ and $8.0$ for the small lattice
are considered to be preliminary and indeed
 we do not use these data for our main studies
by the previous section.}

\begin{figure}[hbt]
\begin{center}
\scalebox{0.3}{
\includegraphics{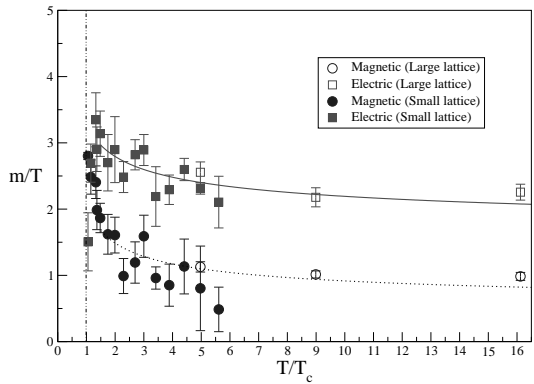} }
\caption{The temperature dependence 
including higher temperature points on the 
large lattice $ 32^2 \times 48 \times 6 $.  
\label{emt2} }
\end{center}
\end{figure}

Although it is very difficult to measure the long distance gluon propagators,
we can add the higher temperature results ( $T/T_c=8.99, 16.12$ )
summarized in Table \ref{mhtr}.
In Fig. \ref{emt2} we again fit the data including these new points.
The fit for the large lattice data by Eq. (\ref{scale}) hence results in
\begin{equation}
\begin{array}{clc}
C_e &= 1.69(4),   &\chi^2/NDF = 0.66,\\ 
C_m &= 0.549(16), &\chi^2/NDF = 1.27 
\end{array}
\end{equation}
These results are shown in Fig. \ref{emt2}.
It should be noticed that 
$C_e$ is the same value as given in Eq. (\ref{fitdata1}),
 while $C_m$ $\sim 10 \%$ larger on the large lattice and
is very close to $m_g=0.568 g^2 T$ calculated
by the self-consistent inclusion technique
in Ref. \cite{Alexanian}.

\begin{table}[t]
\caption{
Data extracted from the small and large lattices
in the same coupling regions, $\beta=7.5$ and $\beta=8.0$.
\label{hight}}
\vspace{0.25cm}
\begin{center}
\renewcommand{\arraystretch}{1.00}
\begin{tabular}{llcccc}
\hline
\hline
$\beta$ &
$\Delta \tau$
& $m_m a({\bf p}\neq 0) $
& $m_e a({\bf p}\neq 0) $\\
\hline
\multicolumn{4}{c}{small lattice size $20^2\times 32 \times 6$ }\\
\hline
7.5 &0.05            &0.301(15)          &0.455(09)\\
8.0 &0.04            &0.311(09)          &0.457(14)\\
\hline
\multicolumn{4}{c}{large lattice size $32^2\times 48 \times 6$ }\\
\hline
7.5 &0.05           &0.231(03)         &0.433(08)\\
8.0 &0.04           &0.214(04)         &0.406(14)\\
\hline
\hline
\end{tabular}
\end{center}
\end{table}

\begin{table}[t]
\renewcommand{\arraystretch}{1.00}
\caption{
Simulation parameters and screening masses
for the large lattice $32^2 \times 48 \times 6$.
Lattice scales are estimated by Refs. \cite{QCD_TARO,Allton}.
\label{mhtr}}
\vspace{0.25cm}
\begin{center}
\begin{tabular}{lcccccc}
\hline
\hline
$\beta$ &$a^{-1}$[GeV] & T[MeV] & $T/T_c$
& $m_m/T$ & $m_e/T$ \\
\hline
7.0 & 7.64&1274& 4.97&1.128(78)&2.556(156)\\
7.5 &13.8 &2303& 8.99&1.014(54)&2.178(144)\\
8.0 &24.7 &4127&16.12&0.984(60)&2.256(120)\\
\hline
\hline
\end{tabular}
\end{center}
\end{table}

\vspace{4.0cm}

\section{Conclusions}

We have measured the gluon propagators and obtained
the electric and magnetic masses
by lattice QCD simulations in the quenched approximation for SU(3)
between $T=T_c$ and $6T_c$.
Features of the QGP in this temperature region will be extensively studied
theoretically and experimentally in the near future.

Our screening mass studies are the first reliable measurement in $SU(3)$
lattice calculation.
We mainly investigate the temperature dependence for
the electric and magnetic masses which do not vanish on
$20^2\times 32 \times 6 $ lattices.
In all temperature regions we find that the electric mass $m_e$
is always larger than the magnetic one $m_m$,  
except near critical temperature point.
As the temperature goes down toward $T_c$, $m_e/T$ drops down quickly,
while $m_m/T$ is still going up.
Consequently, using data above $T/T_c \sim 1.5$ we conclude that
 the scalings $m_e\sim gT$ and $m_m\sim g^2T$ work well.
Furthermore, 
a HTL resummation calculation has recently been developed and
compared with nonperturbative lattice simulations.
We have also compared our numerical results
with LOP and HTL resummation and 
find a good improvement of the HTL electric mass.
These comparison studies of $SU(3)$ screening masses
qualitatively seem to agree with the case of $SU(2)$ \cite{Heller}.

The electric masses obtained here are not consistent with those obtained by
 heavy $q\bar{q}$ potential calculations from an $SU(3)$ Polyakov loop 
correlator at finite temperature in Refs. \cite{Gao,Kaczmarek}.
In Ref. \cite{Kaczmarek}, the authors did extensive analyses
with three different temporal extents and two different gauge actions,
obtaining a very reliable potential as a function of the temperature.
They observe that the potential above $T_c$ cannot be described 
properly by the leading order perturbation calculation up to a few
$T_c$: They exclude the two-gluon exchange as the dominant screening
mechanism, and suggest that some kind of one-gluon exchange may
describe the potential effectively as a result of the complex 
interaction, and that at about $(1.5-3) T_c$ a mixture of
one- and two-gluon exchange may explain the behavior. 
Therefore, due to the ambiguity of the fitting assumptions, it is not
clear whether we can compare our screening masses directly with 
those obtained by the potential calculation.

In order to investigate the nature of the QGP, especially the excitation
modes in the plasma,
Datta and Gupta recently calculated glueball masses at finite
temperature and made an interesting observation.
They measured the screening masses of $A_1^{++}$ (scalar) and
 $A_2^{--}$ (glueball),
which allow two- and three-gluon exchange,
and their ratio $\sim 1.7$ is near 3/2.
The $A_2^{--}$ mass is twice that obtained by Kaczmarek {\it et al}, and
shows similar temperature dependence.  
There are now several nonperturbative methods to study QGP:
our direct measurement of the gluon propagators, 
glueball screening masses, and Polyakov line correlators.
These analyses strongly suggest that the QGP above $T_c$ is far from 
a free gas and has a nontrivial structure. 
Much more detailed analyses in future are highly desirable.

The screening mass on the lattice is extracted
from the gauge dependent propagator,
and the magnetic mass is not well defined in perturbation
theory. We have nonperturbatively confirmed the gauge invariance 
of both screening masses.
In Ref. \cite{Cucchieri2}
it was reported that the $SU(2)$ magnetic propagator
exhibits a complicated gauge dependent structure at low momentum.
Therefore, since the gauge dependence for the screening masses 
is investigated within Lorentz-type gauge fixing 
based on stochastic gauge quantization in this study, we 
plan to extend our analysis to a simulation with Coulomb-type gauge fixing.

We have seen a qualitative difference of the gluon dynamics
between the confinement and deconfinement phase
by direct propagator measurement.
The electric and magnetic gluons in the confinement phase
indicate a very massive particle behavior,
while after the phase transition, they have a finite mass.
In addition, in the deconfinement phase, 
the magnetic gluon at short distances 
seems to be still in the confinement phase.
This may be related to the difficulty 
of the perturbative argument for spatial gluon components
and the fact that a magnetic Wilson loop gives non-zero spatial string tension  
even at high temperatures \cite{Fingberg}.   

The magnetic mass has been the subject of many discussions. 
Perturbatively, it is difficult to handle. 
To our knowledge, there is no complete
perturbative calculation which is free from any assumption or model.  
The naive expectation is that it vanishes, but it is necessary to have
a finite value as a cutoff factor in the infrared regime.
On the other hand, the finite spatial string tension even at $T>T_c$
indicates ``confinement" in the magnetic sector. 

Our screening gluon magnetic propagators indeed show nontrivial behavior.
At very short distances, this may be consistent with massless behavior,
but at finite distance we cannot fit them by a simple ansatz.
The behavior is distance dependent.
Confinement is a long range property, and the propagators there drop. 
Therefore it will be a very interesting
task in future to investigate the magnetic propagators at very long distance.   

We calculated the gluon propagators 
on the larger lattice $32^2\times 48\times 6 $ 
and observed a large error and strange behavior at long distances.
Nevertheless, the screening masses were estimated, 
and we find that the magnetic mass is sensitive to the lattice size effect. 
Thus on a too small lattice we cannot deal 
with the magnetic mass consistently. 
Moreover, the simulations even at higher temperature $T/T_c\sim 9$ and $16$ 
show that nonperturbative results are far from LOP.
This observation is compatible with that of Ref. \cite{Kajantie,Heller}.

For quantization with gauge fixing, stochastic gauge fixing was adopted.
We think the stochastic gauge fixing has better
features to reduce some of the difficulties of nonperturbative gauge fixing.
 It is consequently possible 
to do a practical simulation of gluon screenings effectively.
However, we also see that the gluon propagators have large fluctuations and
unexpected behavior at long distances 
and may need further calculations.

The color screening data we obtained here are useful information
for QGP phenomenology, for instance, jet quenching or the heavy quark potential.
We plan to study the $q\bar{q}$ as well as the $qq$ potential 
relating a baryon bound state,
a nonperturbative QCD vertex calculation, 
quark propagators, etc., using stochastic gauge fixing,
which will help us to understand QGP.

\section{Acknowledgement} 
\

We would like to thank A. Ni$\acute{\mbox{e}}$gawa,
S. Muroya, T. Inagaki and I. Pushkina for many
helpful discussions. 
The calculation was done on an SX-5(NEC) vector-parallel computer.
We really appreciate the warm hospitality and support
of the RCNP administrators. 
An HPC computer at INSAM, Hiroshima University was also used.
This work was supported by Grant-in-Aids for Scientific Research by
Monbu-Kagaku-sho (No.11440080, No. 12554008, and No. 13135216)

\end{document}